\documentclass[useAMS,usenatbib]{mn2e}
\usepackage{Times}

\usepackage[T1]{fontenc} 
\usepackage{ae,aecompl}
\usepackage{amsmath}
\usepackage{amsfonts}
\usepackage{amssymb}
\usepackage{graphicx}
\usepackage{natbib}

\makeatletter
\newcommand\footnoteref[1]{\protected@xdef\@thefnmark{\ref{#1}}\@footnotemark}
\makeatother

\title[IX~Draconis -- a curious ER~UMa-type dwarf nova]{IX~Draconis -- a curious ER~UMa-type dwarf nova}

\author[M.~Otulakowska-Hypka, A.~Olech et al.]
{M.~Otulakowska-Hypka$^{1}$\thanks{E-mail:magdaot@camk.edu.pl}, A.~Olech$^{1}$, E.~de~Miguel$^{2}$, A.~Rutkowski$^{3,4}$, R.~Koff$^{5}$, K.~B\k{a}kowska$^{1,6}$\\
$^{1}$N. Copernicus Astronomical Center, Polish Academy of Sciences, ul. Bartycka 18, 00-716 Warsaw, Poland\\
$^{2}$Departamento de Fisica Aplicada, Facultad de Ciencias Experimentales, Universidad de Huelva, 21071 Huelva, Spain\\
$^{3}$TUBITAK National Observatory, Akdeniz University Campus, 07058 Antalya, Turkey\\
$^{4}$Astronomical Observatory, Jagiellonian University, ul. Orla 171, 30-244 Krak\'ow, Poland\\
$^{5}$Center for Backyard Astrophysics, Antelope Hills Observatory, 980 Antelope Drive West, Bennett, CO 80102, USA\\
$^{6}$Astronomical Observatory Institute, Faculty of Physics, A. Mickiewicz University, S\l oneczna 36, 60-286 Pozna\'n, Poland}

\begin{document}

\date{Accepted YYYY Month DD. Received YYYY Month DD; in original form YYYY Month DD}

\pagerange{\pageref{firstpage}--\pageref{lastpage}} \pubyear{YYYY}

\maketitle

\label{firstpage}

\begin{abstract}
We report results of an extensive world-wide observing campaign devoted to a very active dwarf nova star -- IX~Draconis.
We investigated photometric behaviour of the system to derive its basic outburst properties and understand peculiarities of IX~Draconis as well as other active cataclysmic variables, in particular dwarf novae of the ER~UMa-type.
In order to measure fundamental parameters of the system, we carried out analyses of the light curve, $O-C$ diagram, and power spectra.
During over two months of observations we detected two superoutbursts and several normal outbursts. 
The \textit{V} magnitude of the star varied in the range $14.6-18.2~\rm{mag}$. 
Superoutbursts occur regularly with the supercycle length ($P_{sc}$) of $58.5\pm0.5~\rm{d}$. 
When analysing data over the past 20 years, we
found that $P_{sc}$ is increasing at a rate of $\dot{P} = 1.8 \times 10^{-3}$.
Normal outbursts appear to be irregular, with typical occurrence times
in the range $3.1 - 4.1~\rm{d}$. 
We detected a double-peaked structure of superhumps during superoutburst, with the secondary maximum becoming dominant near the end of the superoutburst.  
The mean superhump period observed during superoutbursts is $P_{sh} = 0.066982(36)~\rm{d}$ ($96.45\pm0.05~\rm{min}$), which is constant over the last two decades of observations.
Based on the power spectrum analysis, the evaluation of the orbital period was problematic. 
We found two possible values: the first one, $0.06641(3)~\rm{d}$ ($95.63\pm0.04~\rm{min}$), which is in agreement with previous studies and our $O-C$ analysis ($0.06646(2)~\rm{d}$, $95.70\pm0.03~\rm{min}$), and the second one, $0.06482(3)~\rm{d}$ ($93.34\pm0.04~\rm{min}$), which is less likely. 
The evolutionary status of the object depends dramatically on the choice between these two values. 
A spectroscopic determination of the orbital period is needed.
We updated available information on ER~UMa-type stars and present a new set of their basic statistics. 
Thereby, we provide evidence that this class of stars is not uniform.
\end{abstract}


\section{Introduction}

Cataclysmic variable (CV) stars are close binary systems which consist of two component stars: a donor (secondary) star which fills its Roche lobe and loses its mass through the inner Lagrangian point onto a white dwarf (primary star). 
Dwarf novae are one of the non-magnetic types of CV systems. They have a white dwarf as a primary, and typically a low-mass main sequence star as secondary. The magnetic field of the white dwarf is weak enough ($<10^6~\rm{G}$), so that the material from the secondary is transported onto the primary through an accretion disk around the white dwarf.
Such a disk is the main source of optical radiation of the system \citep{2001Hellier}.
Mass transfer in dwarf novae leads to outbursts which are explained by the thermal-tidal instability model (TTI) as caused by a sudden gravitational energy release due to accretion of material onto the white dwarf \citep{Osaki1974}.

One of the types of dwarf novae are SU~UMa stars which are characterized by short orbital periods ($P_{orb} < 2.5 ~\rm{h}$). 
Their most important feature is that, beside normal outbursts, they show superoutbursts about $1~\rm{mag}$ brighter and about ten times less frequent 
than normal outbursts. 
During superoutbursts there are observed so-called superhumps, i.e. additional "tooth-shape" modulations in the light curve with amplitudes of a fraction of a magnitude, and periods larger by about $3\%$ than the corresponding
orbital period.

The most active among them are ER~UMa-type stars \citep{1995Kato}. These are objects with extremely short supercycle lengths of about $20~-~60~\rm{d}$, normal outburst cycle lengths of about $3~-~4~\rm{d}$ and amplitudes of outbursts of about three magnitudes, which is small in comparison to typical SU~UMa stars. 
They seem to be standard SU~UMa stars with only a larger activity and luminosity due to their higher mass transfer rates \citep{1996Osaki}. 

The length of the supercycle and the number of normal outbursts in between superoutbursts are specific for each SU~UMa-type star. This feature can be explained by the TTI model \citep{1996Osaki}.
According to this model, normal outbursts remove less material from the disk than was accreted since the last outbursts. Thus, the disk is growing during successive normal outbursts. After it reaches the critical radius ($r \sim 0.46 a$, with $a$ being the binary separation), 
the disk becomes eccentric and tidal effects occur. This leads to a superoutburst, during which the disk is shrinking again.
In this scenario the sypercycle length is set by the mass transfer rate in the way that the observed length of the supercycle is inversely proportional to the mass transfer rate. 
According to the standard theory of evolution of CVs, an object evolves from long to short orbital periods and from high ($\sim~10^{-8}~\rm{M_\odot/yr}$) to low ($\sim~10^{-10}~\rm{M_\odot/yr}$) mass transfer rates \citep{2011Knigge}. 
That is why it is very surprising to find extremely active ER~UMa-type stars among the objects with the shortest orbital periods, or even as period bouncers (which should have the lowest mass transfer rate $\sim~10^{-12}~\rm{M_\odot/yr}$). 

Until now there were discovered only five stars belonging to the ER~UMa-type of dwarf novae, among them IX~Draconis (IX~Dra). 
Thus, studying each one of them is of primary importance.

IX~Dra was detected for the first time by \citet{1980Noguchi}, 
and described as a variable object with an UV~excess and a proper motion of $0.016~\rm{arcsec/yr}$ by \citet{1982Noguchi}.
The object was identified as a cataclysmic variable candidate by \citet{1995Klose}, based on the photometric behaviour and the fact that it has an UV~excess. 
IX~Dra was mentioned in the \textit{"Catalog and Atlas of Cataclysmic Variables"} by \citet{1997Downes} as a dwarf nova candidate of the U~Gem class, variable in the magnitude range of $15.0-18.5~\rm{B}$. 
\citet{1999Liu} published a spectrum of IX~Dra obtained at \textit{V} magnitude of $16.3$ which showed a strong blue continuum and emission lines indicating that the spectrum was not taken during quiescence. This was the definite proof of the CV nature of this object. 
A detailed study on IX~Dra was reported by \citet{2001Ishioka}. They performed a few-months-long observational campaign and found that the object belongs to the ER~UMa-type of stars. They were able to estimate the basic observational parameters of the system: 
supercycle length of $53~\rm{d}$, normal outburst cycle length of $3~-~4~\rm{d}$, outburst amplitude of $2.5~\rm{mag}$, and suspected superhump period of $0.067~\rm{d}$. 
Another extensive photometric observational campaign was presented by \citet{2004Olech}. They reported similar estimates of the main observational parameters of IX~Dra: 
supercycle and cycle lengths equal to $54\pm1~\rm{d}$ and $3.1\pm0.1~\rm{d}$, respectively.
Additionally, they estimated the superhump period as $0.066968(17)~\rm{d}$ and found another period of $0.06646(6)~\rm{d}$ in the light curves. If the latter is the orbital period, the value of the superhump period excess would suggest a very low mass ratio of the binary system, which in turn implies that IX~Dra has a brown dwarf as secondary. That would make IX~Dra one of the most evolved dwarf novae.
\citet{2007Ishioka} presented results of infrared spectroscopy which were performed in order to determine the spectral types of secondary stars of a few CV systems including IX~Dra.
For this system, they found typical features of late-type dwarfs, with the best fit to the spectrum corresponding to a template spectrum of M1--M5 dwarfs with a power-law component with an index~$-3.1$.  
What is more, from the relation between the absolute magnitude and the period, they estimated that the distance to the object is $995~\rm{pc}$, and from the template dwarfs of M1--M6 types the expected distance is $1000~-~3000~\rm{pc}$. 
Although \citet{2007Ishioka} consider that the secondary is a more massive 
star, the possibility that the secondary is actually 
a brown dwarf with a large mass-transfer
rate could not be ruled out.
\citet{2007Pretorius} estimated the distance as $d~=~430^{+340}_{-190}~\rm{pc}$
based on the apparent brightness in the infrared. 
These recent facts were our motivation to study this interesting object and to verify its evolutionary status.
It is very important in the context of our understanding of the still poorly studied ER~UMa-type of stars. 

The paper is arranged as follows. 
In Section~\ref{sec-obs} we give information on our observations and data reduction. 
The global light curve of IX~Dra is presented and analysed in Section~\ref{sec-lc}. 
In Section~\ref{sec-sup} we analyse the periodicity of detected superhumps.
Section~\ref{sec-eruma} is dedicated to a wider interpretation of
properties of IX~Dra and the rest of members of the ER~UMa-type class.
We summarize the conclusions of this work in Section \ref{sec-sum}.

\section{Observations and data reduction}
\label{sec-obs}

\begin{table*}
\begin{minipage}{170mm}
\centering
\caption{Our observations of IX~Dra in 2010.}
\begin{tabular}{c|c|c|c}
\hline
Observatory						&	Country			&	Telescope			&	Observer \\
\hline
Warsaw University Observatory		&	Poland			& 0.6~m					&	M.~Otulakowska-Hypka, A.~Olech \\
TUBITAK Observatory					&	Turkey			& 0.4~m, 0.6~m, 1.0~m	& 	A.~Rutkowski			\\
Observatorio del CIECEM			 	& 	Spain			& 0.25~m				&	E. de Miguel			\\
Antelope Hills Observatory		 	&	USA (CO)		& 10''					&	R. Koff					\\
NF/Observatory					&	USA (NM, AZ) 	&	0.6 m, 24''			&	A. W. Neely				\\
Tzec Maun Foundation Observatory &	USA (NM)		&	14'', 16'', 18 cm	&	K. B\k{a}kowska, M. Otulakowska-Hypka \\
\hline
\end{tabular}
\label{tab-obs}
\end{minipage}
\end{table*}

Observations of IX~Dra reported here were obtained between 2010 October 3 and December 13. During this time the dwarf nova showed two superoutbursts (the first one was observed during 15 nights, and the second one during 11 nights) and seven normal outbursts (observed during 34 nights). The observations in total cover 60 nights with 10613 measurements with the average exposure time of about 120~s, depending on the instrument. 

Altogether IX~Dra was monitored by six observers using eleven telescopes with primary 
mirror diameters ranging from $0.25$ to $1.0~\rm{m}$, located in the USA and spread over Europe. Such locations of telescopes allowed us to obtain a beautiful light curve without significant breaks, especially during superoutbursts which are of the greatest importance for us. For much of that time observations were performed almost without breaks, the main limitation came from the weather.
The observers are members of the \textit{CURVE} team \citep{2008Olech, 2009Olech}, see \texttt{http://users.camk.edu.pl/magdaot/curve/}, 
and the \textit{Center for Backyard Astrophysics (CBA)} \citep{2003Patterson, 2005Patterson}, see \texttt{http://cbastro.org/}.

The first of our instruments used during this observational campaign was the $0.6~\rm{m}$ telescope of the Warsaw University, located in Ostrowik, Poland. 
The observations were also carried out at the TUBITAK National Observatory in Turkey with $0.4~\rm{m}$, $0.6~\rm{m}$, and $1.0~\rm{m}$ telescopes. 
IX~Dra was also observed remotely using small telescopes
of the Tzec Maun Foundation, located in New Mexico, USA. 
Additional observations were done with the Webscope ($24''$) and Super Lotis ($0.6~\rm{m}$) telescopes of the NF/~Observatory in New Mexico and Arizona, USA \citep{2008Grauer}.
The input from the \textit{CBA} came from a $0.25~\rm{m}$ telescope at
Observatorio del CIECEM, Huelva, Spain, and a $10''$ telescope at
Antelope Hills Observatory, Colorado, USA. Information on the observations are gathered in Tab.~\ref{tab-obs}.

\begin{figure}
\begin{center}
\includegraphics[width=0.45\textwidth]{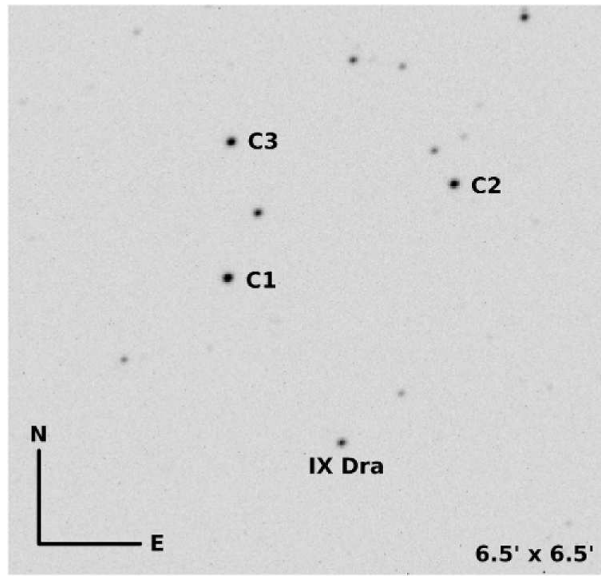}
\caption{Position of IX Dra and three comparison stars on a chart.}
\label{fig-sky}
\end{center}
\end{figure}

The object was observed with the clear filter ("white light") in order to obtain precise photometry of the star in quiescence and also at minimum light -- below $18~\rm{mag}$. 
Flat-field, dark, and bias corrections were performed on the raw files in a standard way.
In the case of the observations with small telescopes from \textit{CBA}, 
this was done with commercially available software. For the rest of the 
observations, standard IRAF routines\footnote{IRAF is distributed by the National Optical Astronomy Observatory, which is operated by the Association of Universities for Research in Astronomy, Inc., under a cooperative agreement with the National Science Foundation.} 
were used.
To extract the profile photometry we used DAOphot~II and Allstar packages \citep{1987Stetson}. 
Relative unfiltered magnitudes of IX~Dra were derived from the difference between the magnitude of the object and the mean magnitude of three comparison stars which are shown in Fig.~\ref{fig-sky}.

Instrumental magnitudes were transformed into the standard \textit{V} system using photometric data from \citet{Henden1995}.
Such a transformed global light curve of IX~Dra spanning over two months of observations
is presented and analysed in the next section.

Typical photometry errors of our measurements
are in the range $0.007-0.21~\rm{mag}$ during the bright state, 
and $0.01-0.25~\rm{mag}$ at the minimum light. 
The median value of the precision equals to $0.019$ and $0.023~\rm{mag}$, respectively.

\section{Global light curve}
\label{sec-lc}

For convenience, from now on we use only a day number, $\rm{HJD}-2455000~[\rm{d}]$, to refer to our observations.
Our observing campaign started on day $473$.

\begin{figure*}
\centering
\includegraphics[width=0.43\textwidth,angle=270]{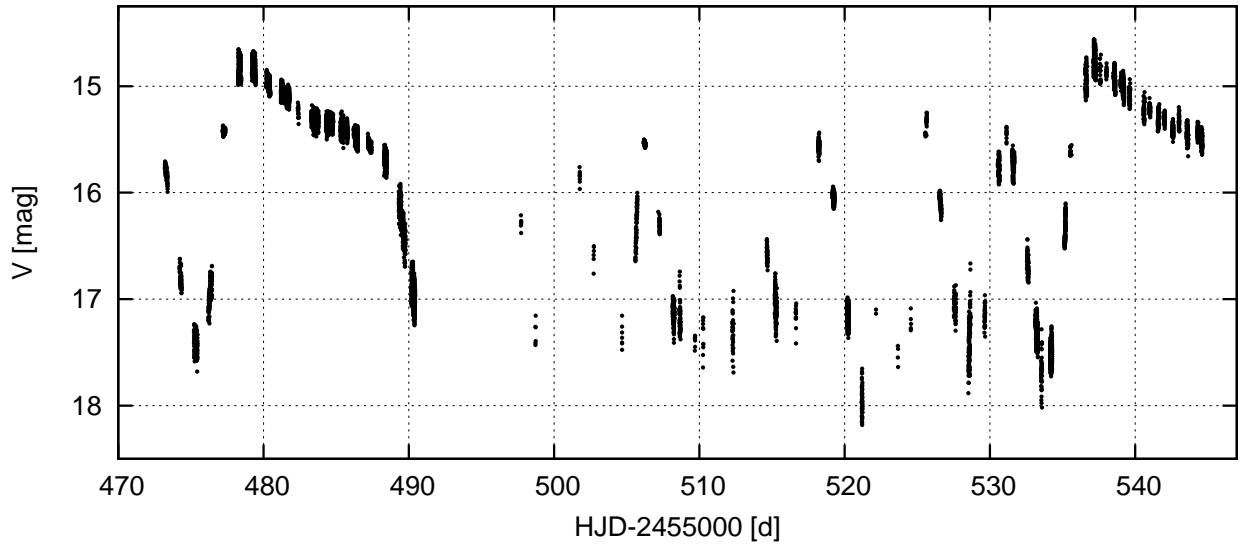}
\caption{Global light curve of IX~Dra during our observational campaign}
\label{fig:lc}
\end{figure*}

Figure~\ref{fig:lc} presents the light curve of IX~Dra during the whole campaign. We were lucky to start the observations at the time of appearance of the precursor of the first of two superoutbursts. 
On day $476$ the superoutburst started and lasted for $15~\rm{days}$ ($2~\rm{days}$ of initial rise, $11~\rm{days}$ of the plateau phase, and $2~\rm{days}$ of final decline). The coverage of this part of the light curve turned out to be the best of the campaign, 
as we were able to collect data during every single night from at least one of the observatories. 
This superoutburst was followed by a set of normal outburst which in total lasted for $43~\rm{days}$. We were only able to observe IX~Dra on 31 of these nights, due to adverse weather conditions.
On day $534$ the second superoutburst started and was observed over the
following $10~\rm{days}$. Weather conditions were not as good as during the first superoutburst.

The most immediate conclusion from the light curve is probably the fact that IX~Dra is an extremely active dwarf nova. Every outburst is almost at once followed by another one. There is no evident quiescence period visible in the light curve between these outbursts. This photometric behaviour is typical for dwarf novae stars of the ER~UMa-type \citep{1996Osaki}.

During superoutbursts the brightness reaches $V~\approx 14.7~\rm{mag}$ at maximum, and fades to $V~\approx 18~\rm{mag}$ at minimum, resulting in the superoutburst amplitude of $A_{s}\approx3.3~\rm{mag}$. 
In turn during normal outbursts the brightness range spreads from $V~\approx 15.3~\rm{mag}$ at maximum to $V~\approx 18.1~\rm{mag}$ at minimum, giving the amplitude of $A_{n}\approx2.8~\rm{mag}$.
These values are typical for ER~UMa stars, see Tab.~\ref{tab:eruma}.

\subsection{Supercycle length}

Until now there are only three previously reported measurements of the supercycle length ($P_{sc}$) for IX~Dra.
The first one, $P_{sc}^1=45.7~\rm{d}$, was given by \citet{1995Klose} with no reported estimate of the corresponding uncertainty. 
The author had very unevenly sampled data, so the periodicity analysis was challenging. 
This uncertainty, however, can be significant, as the author claimed that the period $P_{sc}^1$ did not match all his data. 
Also the second measurment, $P_{sc}^2=53~\rm{d}$, was presented by \citet{2001Ishioka} with no uncertainties.
In this case authors did not give a reason for that. 
The third measurement of the supercycle length, $P_{sc}^3=54\pm1~\rm{d}$, was presented by \citet{2004Olech}.
Considering the probably large uncertainty of the first of these values, 
the results of \citet{2001Ishioka} and \citet{2004Olech} seemed to suggest
a constant value of $P_{sc}$ for IX~Dra.

To derive $P_{sc}$ from our observations, we computed the power spectrum using the ANOVA code of \citet{1996Schwarzenberg-Czerny} only for the part of the light curve which covers both superoutbursts. 
We found that the most prominent peak of the power spectrum is the one at the frequency $0.0171(2)~\rm{c/d}$ \textit{i.e.} $P_{sc}=58.5\pm0.5~\rm{d}$. 

\begin{figure}
\begin{center}
\includegraphics[width=0.34\textwidth,angle=270]{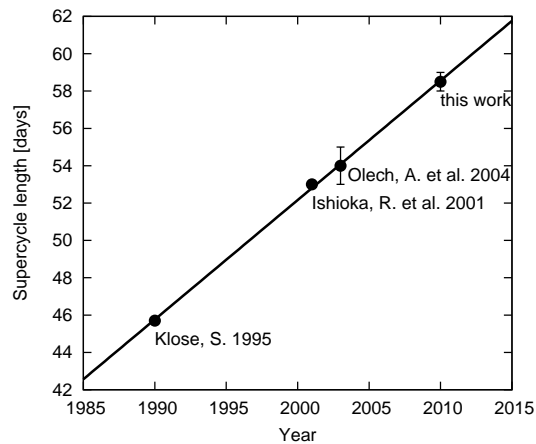}
\caption{
The increasing supercycle length of IX~Dra during the past twenty years. 
The data points are taken from the literature. 
Uncertainties are given when available.
The line represents the best fit to the data.
The corresponding rate of increase of the period is $\dot{P} = 1.8 \times 10^{-3}$.}
\label{fig:Tsc}
\end{center}
\end{figure}

Thus, we provide a new estimate of the supercycle length for this object and thereby we show that $P_{sc}$ has increased during the past twenty years (see Fig.~\ref{fig:Tsc}). The corresponding rate of increase of the period is $\dot{P} = 1.8 \times 10^{-3}$.

In spite of the fact that our estimate of $P_{sc}$ is based on only two observed superoutbursts, such a change of the supercycle length seems reasonable, because such changes were also observed for other ER~UMa-type stars.
For instance \citet{2001Kato} reported an extremely variable supercycle length of V1159~Ori. In his Fig.~2 one can see rapidly changing jumps of this value between $44.6$ and $53.3~\rm{d}$ over a span of about five years. 
RZ~LMi has a supercycle length decreasing with the rate of $\dot{P} = -1.7 \times 10^{-3}$, in turn ER~UMa has the supercycle length increasing with the rate $\dot{P} \simeq 4 \times 10^{-3}$ \citep{1995Robertson}.

\begin{figure}
\centering
\includegraphics[width=0.27\textwidth,angle=270]{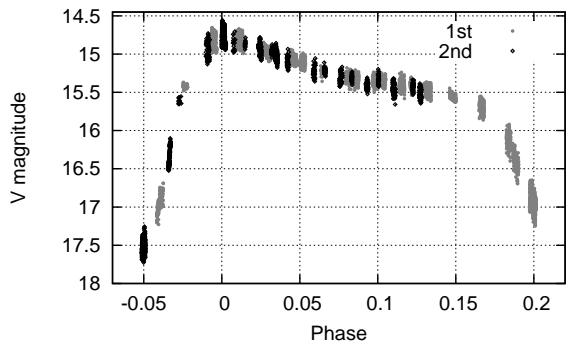}
\caption{Light curve of IX~Dra during superoutbursts obtained by folding the light curve with $P_{sc}=58.5\pm0.5~\rm{d}$. Grey circles indicate the data of the first superoutburst, black diamonds represent the data of the second superoutburst.}
\label{fig:lcfolded}
\end{figure}

We used the $P_{sc}$ to create the folded light curve of our both superoutbursts. It is shown in Fig.~\ref{fig:lcfolded}.

Although the light curve folded with the $P_{sc}$ looks reasonable, and despite the fact that significant changes of $P_{sc}$ are apparently typical for ER~UMa-type stars, we decided to investigate our result more precisely to make sure that it is not an incidental effect based on only two perhaps untypical superoutbursts.
We looked for additional recent photometric data of the object, which cover a greater time range.
In the archive of \textit{The American Association of Variable Star Observers (AAVSO)} we found a light curve for IX~Dra. 
Its time coverage was quite poor, probably because of the fact that the object reaches 18~mag in quiescence and thus it is then too faint for amateur observers. 
However, we were lucky to find one superoutburst with sufficient time coverage which was just following our observational run.
\begin{figure*}
\centering
\includegraphics[width=0.35\textwidth,angle=270]{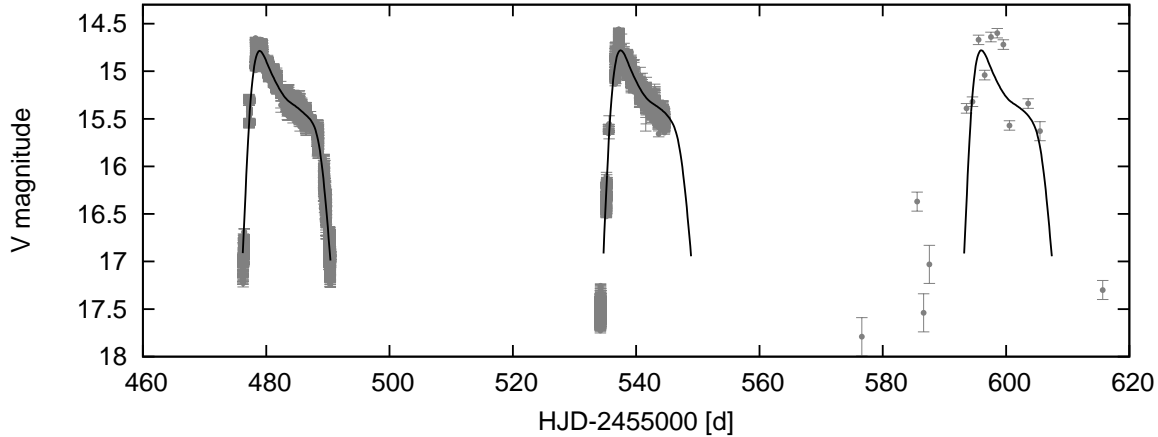}
\caption{Observed superouturst of IX~Dra (points) from our observational campaign and the succeeding superoutburst from the \textit{AAVSO} archive, suggesting that the $P_{sc}=58.5~\rm{d}$ is consistent with all three superoutbursts. 
Lines indicate anticipated occurrences of superoutbursts with the supercycle length $P_{sc}$.}
\label{fig:aavso}
\end{figure*}
In Fig.~\ref{fig:aavso} we present both superoutbursts observed during our campaign together with the succeeding superoutbursts from the \textit{AAVSO} archive (at day $\sim600$) which is in agreement with our supercycle estimate $P_{sc}=58.5\pm0.5~\rm{d}$. 
The superoutburst profile, which is shown as the solid line in the Fig.~\ref{fig:aavso}, was obtained by completing an analytical fit using Bezier curves to the first superoutburst. The derived profile was copied forward with supercycle period $P_{sc}$.

There is one more interesting issue possible to check out in the context of the increase of the supercycle length of IX~Dra: the future evolution of the system. 
Assuming that the supercycle length will increase in the same way in the future, we can predict when IX~Dra would reach the status of SU~UMa and WZ~Sge stars.
Typical supercycle lengths for these types of stars are a few hundred days and decades, respectively \citep{2001Hellier}. 
Thus, assuming that this increasing trend of $P_{sc}$ will continue, we calculated that IX~Dra would attain the stage of SU~UMa with the $P_{sc}=300~\rm{d}$ in almost $400~\rm{years}$ from now. 
In turn in about $7000~\rm{years}$ the system would become a WZ~Sge-type star with the supercycle length of $10~\rm{years}$.
We wish to stress that this is only a speculation which is in contradiction 
with the most likely scenario of the evolution of the object \textit{i.e.} 
that IX~Dra has a brown dwarf donor, and that the high mass transfer rate is an unusual state (see \cite{2004Olech} and Sec.~\ref{sec:per_int} for details).

\subsection{Normal cycle length}

\begin{figure}
\begin{center}
\includegraphics[width=0.19\textwidth,angle=270]{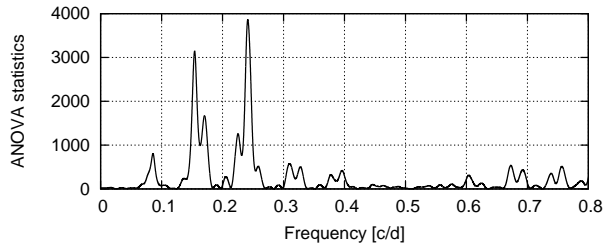}
\caption{Power spectrum for the light curve of normal outbursts with the highest peak corresponding to $P_{c}^1=4.1\pm0.1\rm{d}$.}
\label{fig:per-normal}
\end{center}
\end{figure}

In contrast to the supercycle, there is not such a straight periodicity in the observed occurrences of normal outbursts. 
We computed power spectra for the light curve covering only normal outbursts, using the ANOVA code of \citet{1996Schwarzenberg-Czerny}. 
We ran a number of tests with a variety of frequency ranges and steps but without any fully satisfactory result. 
One can say that the most prominent peak is the one shown in Fig.~\ref{fig:per-normal} and corresponding to the frequency $0.24\pm0.05~\rm{c/d}$ \textit{i.e.} $P_{c}^1=4.1\pm0.1~\rm{d}$. 

\begin{figure*}
\centering
\includegraphics[width=0.35\textwidth,angle=270]{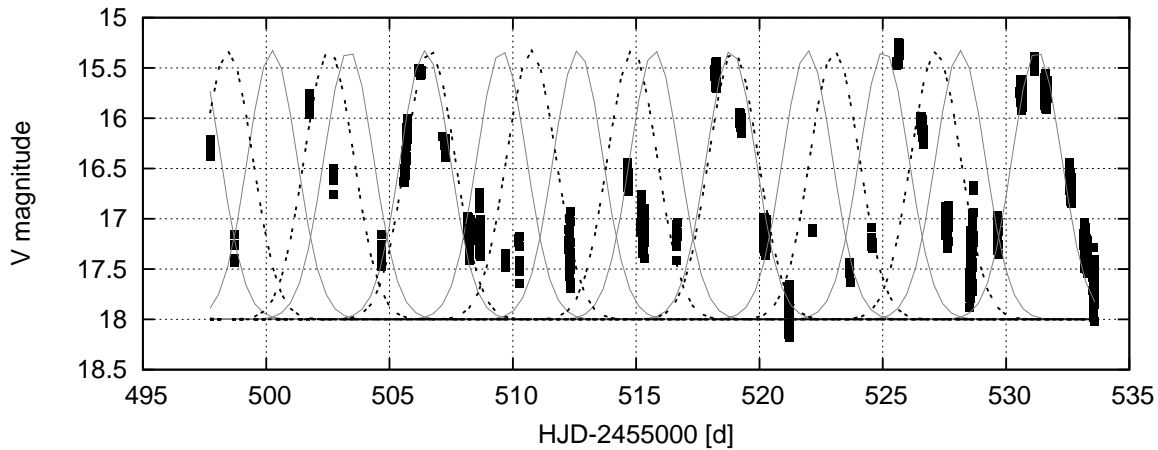}
\caption{
Normal outbursts of IX~Dra during our observational campaign do not appear to be periodic.
We show the observational data (black dots) with two superimposed curves indicating regular anticipated occurrences of normal outbursts with the most likely period estimated from our observations $P_{c}^1=4.1~\rm{d}$ (black dashed line), and with the value of the normal cycle length from the literature, $P_{c}^2=3.1~\rm{d}$ (grey solid line). 
Neither period decribes the lightcurve well.
}
\label{fig:normal}
\end{figure*}

However, we cannot accept this value as the normal cycle length of the system, $P_c$. 
The folded light curve with such a period did not give an adequate result. 
Our conclusion is that normal outbursts observed during our campaign did not occur at regular intervals. 
On the other hand, we know the previously estimated value of the normal cycle length of IX~Dra from \citet{2004Olech}: $P_{c}^2=3.1\pm0.1~\rm{d}$.
To make this issue clear, we plot the light curve with superimposed two curves indicating regular anticipated occurrences of normal outbursts with 
$P_{c}^1$ and another one with $P_{c}^2$.  
First, we made a fit to the last ouburst (at $531~\rm{d}$), which is the only one with a sufficient time coverage, and then repeated both fits backward in time with the given periods. See Fig.~\ref{fig:normal}.

It is now evident that normal outbursts are not exactly periodic, their occurrence seems to be pretty random. Only two of them, observed at days $\sim~506$ and $531$, emerged more or less according to both predictions, $P_{c}^1$ and $P_{c}^2$.

\section{Superhumps}
\label{sec-sup}

\begin{figure*}
\begin{center}
\includegraphics[width=0.8\textwidth]{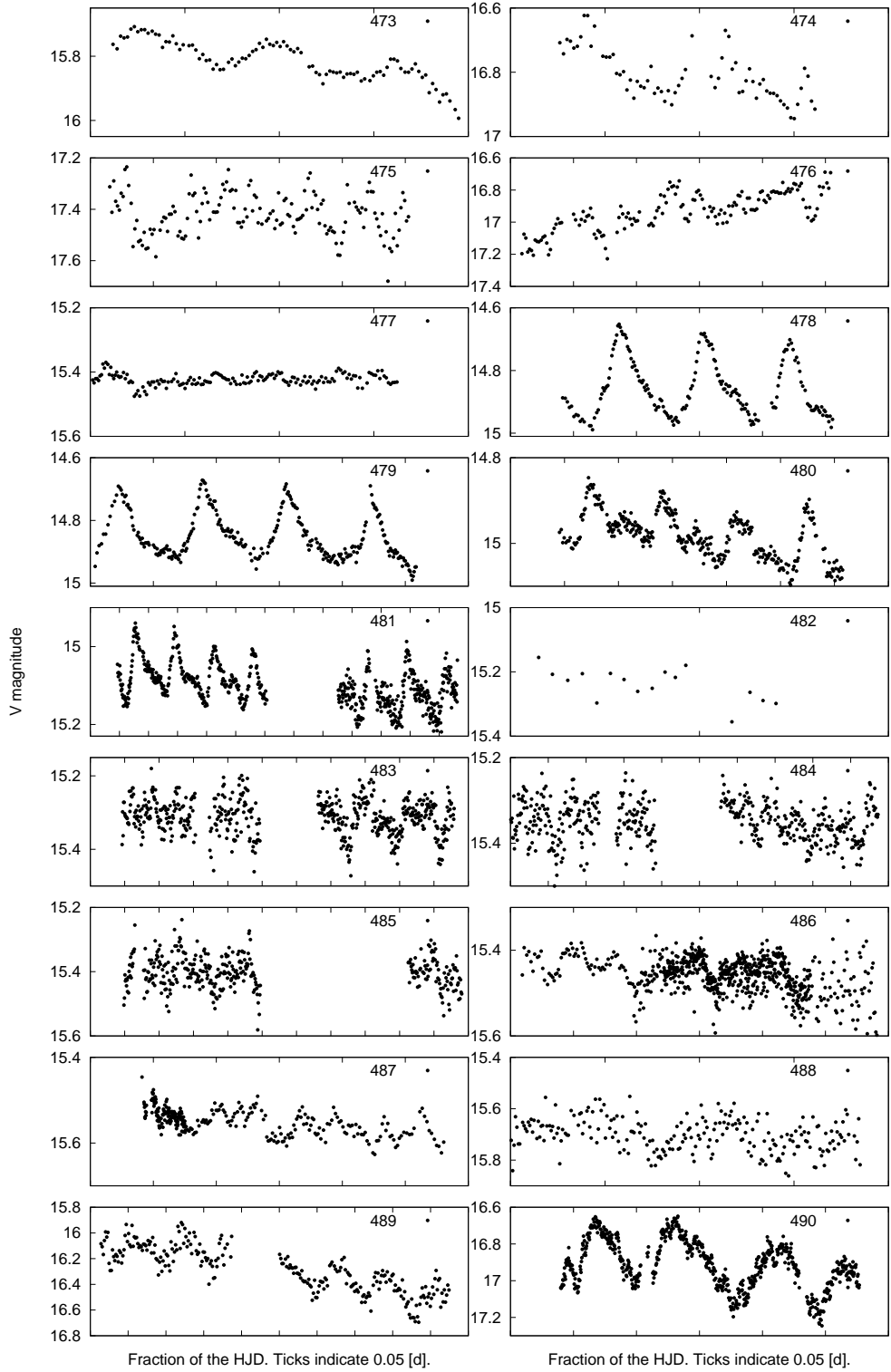}
\caption{Superhumps at each day of the first superoutburst. Consecutive nights are denoted by dates given as HJD-2455000 [d]}
\label{fig:superhumps}
\end{center}
\end{figure*}

The light curve of IX~Dra shows clear superhumps during superoutbursts as well as during normal outbursts. In Fig.~\ref{fig:superhumps} we show nightly curves of the first superoutburst to illustrate their behaviour. 

The first three nights ($473~-~475$) correspond to the precursor and show small amplitude modulations of about $0.15~\rm{mag}$. 
The next two nights ($476~-~477$) correspond to the quick initial phase of the superoutbursts and show modulations with the amplitude reaching $0.2~\rm{mag}$. 
Just after hitting the maximum brightness of the light curve (days $478~-~479$), there are visible clear and accurate, so-called "tooth-shape", superhumps with the maximum amplitude $0.35~\rm{mag}$, decreasing with time to $0.24~\rm{mag}$ during the whole plateau phase.
A very interesting behaviour begins at day $480$: aside from the normal superhumps we observe secondary maxima shifted by $0.5$ in phase with respect to primary superhumps, with the amplitude of about $0.05~\rm{mag}$ increasing with time.
After day $482$ the light curve gets complex and even chaotic -- it is quite tricky to tell apart the primary and secondary superhumps. During this time, the amplitude of primary superhumps is decreasing and the amplitude of the secondary humps is increasing. At the end of the superoutburst (day $490$) we see again strong clear humps, with the amplitude reaching $0.4~\rm{mag}$. Although, their shapes differ from the typical "tooth-shape" of classical superhumps. 

To investigate the apparent periodicities in the light curve, we performed power spectra and $O-C$ analyses. They are described in the following subsections.

\subsection{Period analysis}

At first, to refine our period search, we converted the light curve in two ways. For a start, we transformed the magnitudes into intensities to equalize the amplitudes of the superhumps. Then, we detrended the light curve for each day separately, by subtracting a best-fitting parabola. As a result we obtained light curves in units of counts with the average intensity at zero. 

Due to the differing quality of the data of each superoutburst, we decided to analyse them separately. We divided the whole data into three data sets, for: 
\begin{enumerate}
\item the 1st superoutburst,
\item the 2nd superoutburst, and
\item normal outbursts and quiescence.
\end{enumerate}

\begin{figure}
\begin{center}
\includegraphics[width=0.18\textwidth,angle=270]{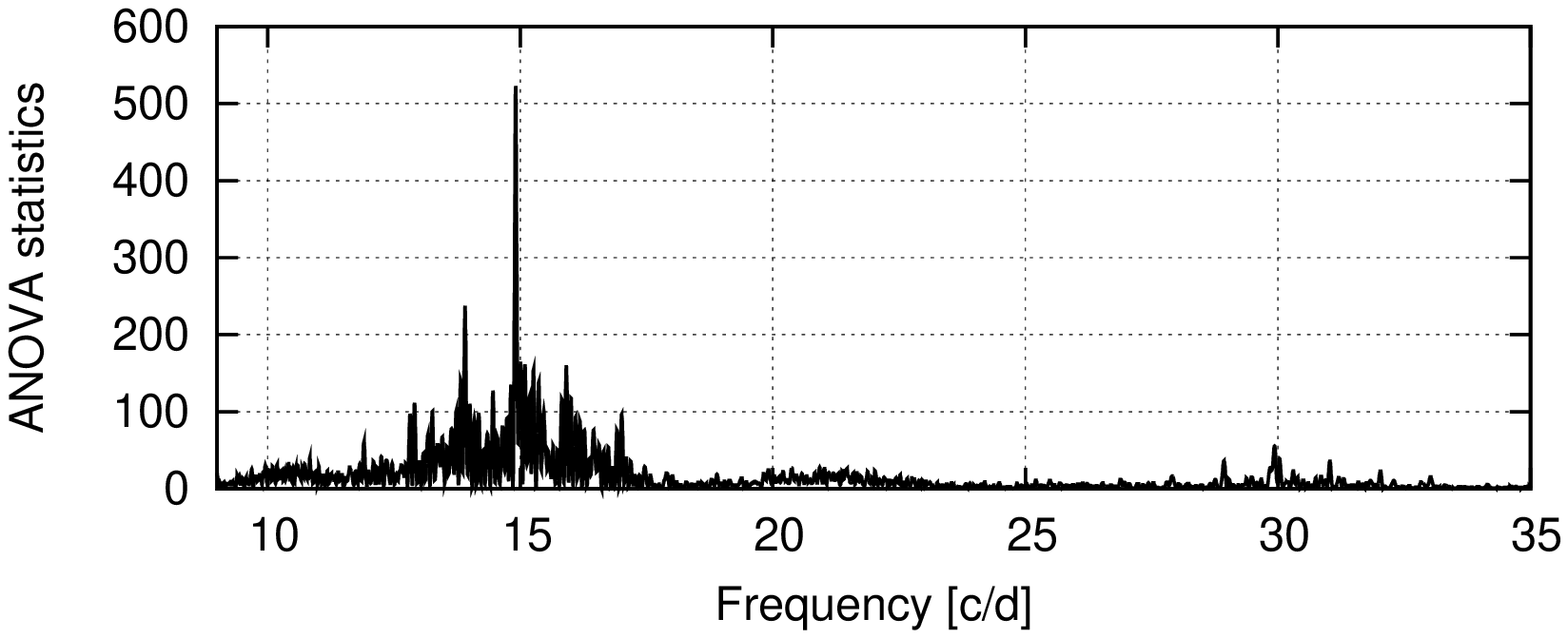}
\includegraphics[width=0.18\textwidth,angle=270]{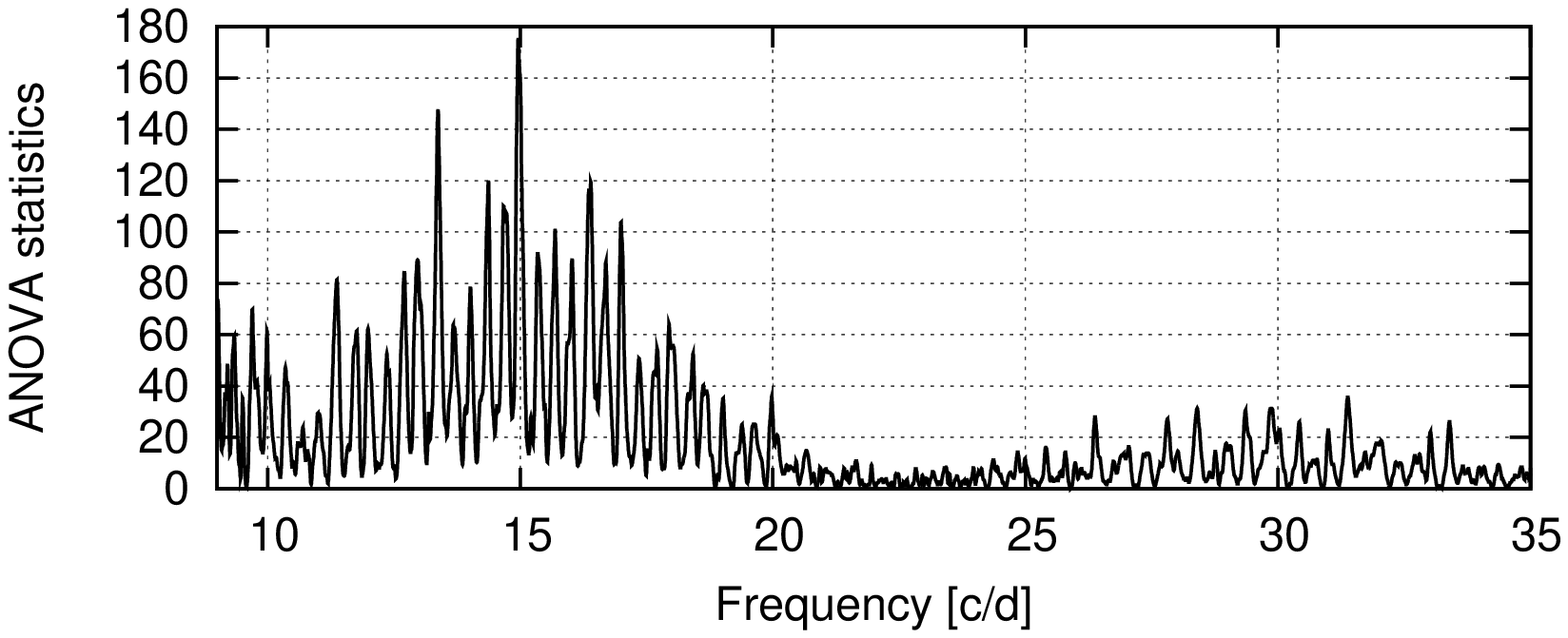}
\includegraphics[width=0.18\textwidth,angle=270]{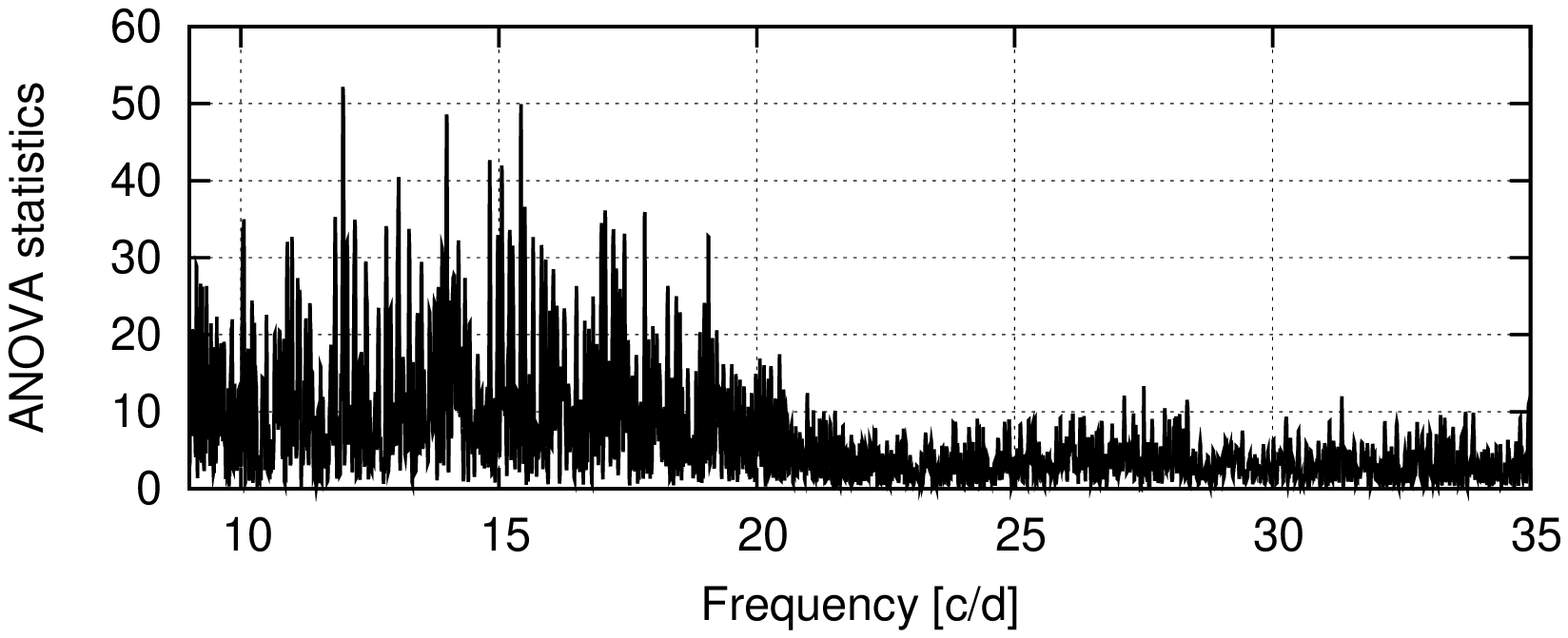}
\caption{Power spectra for the light curve (from top to bottom) of the 1st superoutburst, the 2nd superoutburst, normal outbursts and quiescence.}
\label{fig:per-all}
\end{center}
\end{figure}

For each of these subsets, we performed the period analysis with the ANOVA code of \citet{1996Schwarzenberg-Czerny} to calculate power spectra. 
Results are presented in Fig.~\ref{fig:per-all}. 

\subsubsection{Superoutbursts}

For both superoutbursts there is clearly visible one prominent peak corresponding to the superhump period, $P_{sh}$.
For the first superoutburst it is located 
at the frequency $14.910(7)~\rm{c/d}$ which determines the period $P_{sh}^1=0.06707(3)~\rm{d}$, 
and for the second one the main peak lies 
at the frequency $14.96(2)~\rm{c/d}$, \textit{i.e.} $P_{sh}^2=0.06683(9)~\rm{d}$. 
These period uncertainties are $1$ sigma uncertainties.

We adopt $P_{sh}^1$ as the subsequent superhump period. It is more reliable than $P_{sh}^2$ due to a greater time 
coverage of observations of the first superoutburst and consequently much better S/N ratio of its periodogram. 
We can say that the result from the second superoutburst, taking into account its worse quality, agrees with $P_{sh}^1$ to within $2$ sigma.

\begin{figure}
\begin{center}
\includegraphics[width=0.5\textwidth]{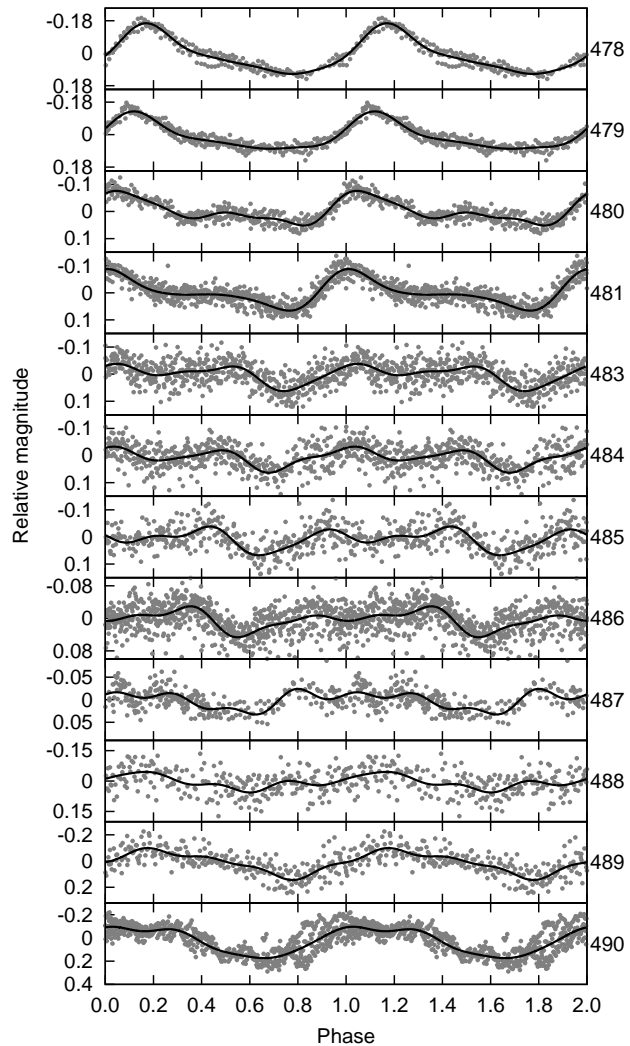}
\caption{
Evolution of superhumps during the first superoutburst. 
Plots show the detrended light curve (grey points), phased with the superhumps period, $P_{sh}^1$, for each day separately. 
Numbers on the right indicate dates of observations (HJD-2455000 [d]). 
The black curve represents our fit to the light curve. See the text for more details.}
\label{fig:shevolution}
\end{center}
\end{figure}

We now consider the analysis of the stability of the phase of superhumps. 
The question about the presence of a phase reversal in ER~UMa-type stars was taken up by a number of researchers, e.g. \citet{2003Kato} and \citet{2004Olech}. 
To check the stability of the phase of superhumps, at first we removed the general decreasing trend from the light curves, and then phased them with the superhump period for each of the nights separately. 
In Fig.~\ref{fig:shevolution} there is such a plot for the data of the first superoutburst. 
For each of the light curves we also plot a corresponding fit to the phased data. 

It is now clearly visible that during the first two nights we observed only the primary superhumps with maxima at phase $\sim1$.
On day $480$ we detected additional humps in the light curve which begins to appear at the phase $\sim0.5$. 
It is present as from now, and its amplitude is continuously increasing over the next days.
On day $485$ we can see that the amplitude of the secondary humps gets higher than the amplitude of the primary superhumps, which is gradually decreasing over the time.
The phase of the secondary humps is also changing -- it is decreasing from $\sim0.5$ to $\sim0.1$. 
Finally, at the day $490$, the place of the primary superhumps is taken by the secondary humps. As was mentioned before, the shape of these final humps differ from the initial "tooth-shape" form. 

\begin{figure}
\begin{center}
\includegraphics[width=0.19\textwidth,angle=270]{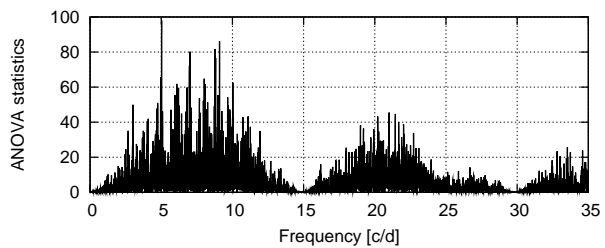}
\caption{Power spectrum for the light curve of the first superoutburst prewhitened with $P_{sh}$.}
\label{fig:pre}
\end{center}
\end{figure}

We performed prewhitening of the detrended light curve in order to search for additional periodicities.
At first, from the light curve we removed a modulation with the superhump period, 
which is shown in Fig.~\ref{fig:shevolution} as a black curve. 
On such data we performed the power spectrum analysis. The result is presented in Fig.~\ref{fig:pre}. 

The power spectrum is pretty noisy, although one could claim that there is some signal around the frequency $\sim9~\rm{c/d}$, and probably its first harmonic at $\sim20~\rm{c/d}$. 
However, this would be unexpected and unlikely.
We detected no trace of orbital modulation during the superoutburst. As reported by \cite{2004Olech}, the orbital signal for IX~Dra was observed during the superoutburst in September 2003. Thus, we also expected to detect it at the frequency $\sim15.05~\rm{c/d}$.

\subsubsection{Normal outbursts and quiescence}
\label{sec:PSofNO}

\begin{figure}
\begin{center}
\includegraphics[width=0.19\textwidth,angle=270]{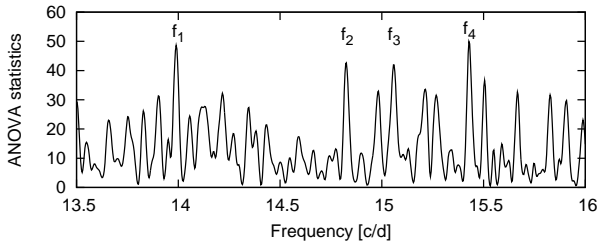}
\caption{Zoom in the power spectrum of the light curve with normal outbursts showing a set of main peaks around $f~=~15~c/d$.}
\label{fig:per-zoom}
\end{center}
\end{figure}

When it comes to the power spectrum of the light curve covering normal outbursts and quiescence, it does not show definite evidence of periodicities (see again Fig.~\ref{fig:per-all}). 
However, one can distinguish a number of peaks with similar power around $\sim15~\rm{c/d}$, corresponding to $P_{sh}^1$ and $P_{sh}^2$.
These most prominent frequencies are labeled as $f_1$, $f_2$, $f_3$, and $f_4$ in Fig.~\ref{fig:per-zoom}, where we show the power 
spectrum zoomed into the region of interest.
The associated periods are presented in Tab.~\ref{tab:freq-periods}.
An interpretation of these signals will be deferred to the following 
sections.

\begin{table}
\centering
\begin{tabular}{c|c}
\hline
frequencies (c/d)	&	periods (d)	\\
\hline
	$f_1 = 13.990(7) $ & $P_{f_1} = 0.07148(4) $ \\
 	$f_2 = 14.826(7) $ & $P_{f_2} = 0.06745(3) $ \\
 	$f_3 = 15.057(7) $ & $P_{f_3} = 0.06641(3) $ \\
 	$f_4 = 15.428(7) $ & $P_{f_4} = 0.06482(3) $ \\
\hline
\end{tabular} 
\caption{The most prominent frequencies for the periodogram of normal outbursts.}
\label{tab:freq-periods}
\end{table}

\subsection{The $O-C$ analysis}

IX~Dra shows clear humps during entire superoutbursts and also throughout normal outbursts and the very short quiescent state. 
In the previous subsection we found a clear periodicity in the data of both superoutbursts, and an indication of the presence of a similar periodicity between them.

Previous studies of this object indicate that apart from typical superhumps, there is an additional light curve modulation observed \citep{2004Olech}.

To perform the $O-C$ analysis of our data, we carefully determined moments of all maxima and, as before, we examined them for each part of the light curve separately. 

\subsubsection{Superoutbursts}

We present results based on only the first superoutburst. This data set is more credible because of its much better quality and full time coverage, in contrast to the second superoutburst.

In this part of the light curve we identified $60$ maxima of superhumps. A linear fit to the data brought the following ephemeris:
\begin{equation}
\label{eq:oc}
\rm{HJD}_{max} = 2455478.2865(5) + 0.066894(6) \times E
\end{equation}

\begin{figure}
\begin{center}
\includegraphics[width=0.62\textwidth,angle=270]{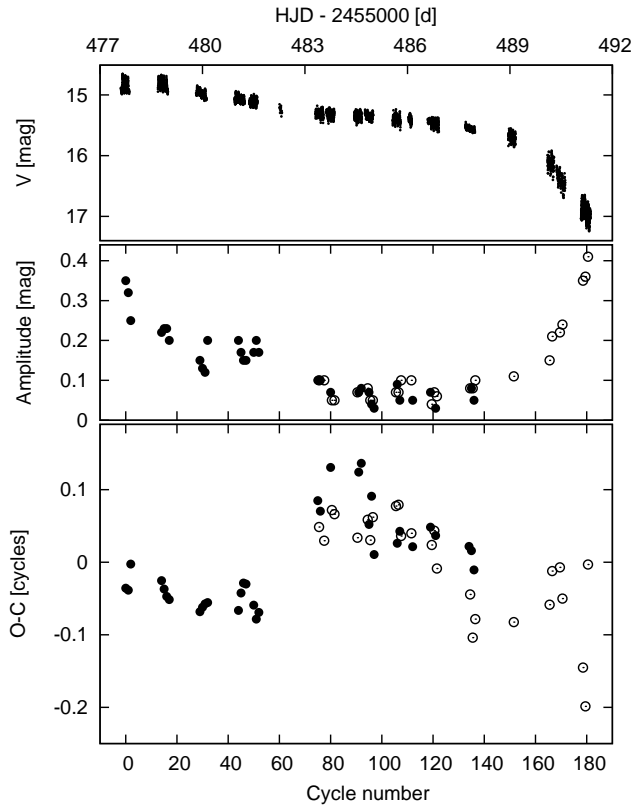}
\caption{
Light curve of the first superoutburst of IX~Dra (top panel), evolution of the amplitude of superhumps (middle panel), and $O-C$ diagram for corresponding superhumps maxima. 
In the two lower plots black dots indicate typical integer values of cycle numbers, open circles present values shifted by a half of a cycle. Details in the text.}
\label{fig:oc}
\end{center}
\end{figure}

The corresponding $O-C$ diagram is presented in Fig.~\ref{fig:oc}.

We list times, errors, cycle numbers, and amplitudes for each superhump in Tab.~\ref{tab:oc}. There are also given the $O-C$ values calculated according to the obtained ephemeris (Eq.~\ref{eq:oc}).

\begin{table}
\centering
\caption{Cycle numbers ($E$), times of maxima ($\rm{T_{max}}$, \textit{i.e.} $\rm{HJD}-2455000~[\rm{d}]$), $O-C$ values, and amplitudes ($A$) of superhumps observed during the first superoutburst.}
\begin{tabular}{c|c|c|c|c}
\hline
$E$		&	$\rm{T_{max}}$	&	error	&	$O-C$ &	$A$	\\
\hline
0.0 & 478.2860 & 0.0013 & -0.0024 	&	0.35	\\
1.0 & 478.3527 & 0.0020 & -0.0026 	&	0.32	\\ 
2.0 & 478.4220 & 0.0016 & -0.0002 	&	0.25	\\ 
14.0 & 479.2231 & 0.0016 & -0.0017 	&	0.22	\\ 
15.0 & 479.2892 & 0.0013 & -0.0025 	&	0.23	\\ 
16.0 & 479.3554 & 0.0014 & -0.0032	&	0.23	\\ 
17.0 & 479.4220 & 0.0020 & -0.0034 	&	0.20	\\ 
29.0 & 480.2235 & 0.0020 & -0.0046 	&	0.15	\\ 
30.0 & 480.2908 & 0.0020 & -0.0042 	&	0.13	\\ 
31.0 & 480.3580 & 0.0020 & -0.0038 	&	0.12	\\ 
32.0 & 480.4250 & 0.0020 & -0.0037 	&	0.20	\\ 
44.0 & 481.2269 & 0.0016 & -0.0045 	&	0.20	\\ 
45.0 & 481.2954 & 0.0017 & -0.0028 	&	0.17	\\ 
46.0 & 481.3632 & 0.0020 & -0.0019 	&	0.15	\\ 
47.0 & 481.4300 & 0.0020 & -0.0020 	&	0.15	\\ 
50.0 & 481.6287 & 0.0025 & -0.0040 	&	0.17	\\ 
51.0 & 481.6943 & 0.0015 & -0.0053 	&	0.20	\\ 
52.0 & 481.7618 & 0.0022 & -0.0046 	&	0.17	\\ 
75.0 & 483.3105 & 0.0050 &  0.0057 	&	0.10	\\ 
75.5 & 483.3415 & 0.0033 &  0.0032 	&	0.10	\\ 
76.0 & 483.3764 & 0.0035 & 0.0047 	&	0.10	\\ 
77.5 & 483.4740 & 0.0040 & 0.0020 	&	0.10	\\ 
80.0 & 483.6480 & 0.0035 & 0.0088 	&	0.07	\\ 
80.5 & 483.6775 & 0.0040 & 0.0048 	&	0.05	\\ 
81.5 & 483.7440 & 0.0040 & 0.0044 	&	0.05	\\ 
90.5 & 484.3438 & 0.0030 & 0.0023 	&	0.07	\\ 
91.0 & 484.3833 & 0.0020 & 0.0083 	&	0.07	\\ 
92.0 & 484.4510 & 0.0020 & 0.0091 	&	0.08	\\ 
94.5 & 484.6130 & 0.0030 & 0.0039 	&	0.08	\\ 
95.0 & 484.6460 & 0.0030 & 0.0035 	&	0.07	\\ 
95.5 & 484.6780 & 0.0040 & 0.0020 	&	0.05	\\ 
96.0 & 484.7155 & 0.0022 & 0.0061 	&	0.04	\\ 
96.5 & 484.7470 & 0.0030 & 0.0042 	&	0.05	\\ 
97.0 & 484.7770 & 0.0030 & 0.0007 	&	0.03	\\ 
105.5 & 485.3500 & 0.0035 & 0.0052 	&	0.07	\\ 
106.0 & 485.3800 & 0.0020 & 0.0017 	&	0.09	\\ 
106.5 & 485.4170 & 0.0040 & 0.0053 	&	0.07	\\ 
107.0 & 485.4480 & 0.0040 & 0.0029 	&	0.05	\\ 
107.5 & 485.4810 & 0.0025 & 0.0024 	&	0.10	\\ 
111.5 & 485.7488 & 0.0040 & 0.0027 	&	0.10	\\ 
112.0 & 485.7810 & 0.0040 & 0.0014 	&	0.05	\\ 
119.0 & 486.2510 & 0.0030 & 0.0032 	&	0.07	\\ 
119.5 & 486.2828 & 0.0035 & 0.0016 	&	0.04	\\ 
120.5 & 486.3510 & 0.0035 & 0.0029 	&	0.07	\\ 
121.0 & 486.3840 & 0.0040 & 0.0025 	&	0.03	\\ 
121.5 & 486.4144 & 0.0040 & -0.0006 	&	0.06	\\ 
134.0 & 487.2525 & 0.0035 & 0.0015	&	0.05	\\ 
134.5 & 487.2815 & 0.0030 & -0.0030 	&	0.08	\\ 
135.0 & 487.3190 & 0.0027 & 0.0011 	&	0.08	\\ 
135.5 & 487.3444 & 0.0020 & -0.0070 	&	0.08	\\ 
136.0 & 487.3841 & 0.0025 & -0.0007 	&	0.05	\\ 
136.5 & 487.4130 & 0.0020 & -0.0053 	&	0.10	\\ 
151.5 & 488.4160 & 0.0050 & -0.0055 	&	0.11	\\ 
165.5 & 489.3540 & 0.0040 & -0.0039 	&	0.15	\\ 
166.5 & 489.4240 & 0.0030 & -0.0008 	&	0.21	\\ 
169.5 & 489.6250 & 0.0030 & -0.0005 	&	0.22	\\ 
170.5 & 489.6890 & 0.0033 & -0.0034 	&	0.24	\\ 
178.5 & 490.2177 & 0.0020 & -0.0097 	&	0.35	\\ 
179.5 & 490.2810 & 0.0030 & -0.0133 	&	0.36	\\ 
180.5 & 490.3610 & 0.0035 & -0.0002  	&	0.41	\\
\hline
\end{tabular}
\label{tab:oc}
\end{table}

There are two very intriguing things about this $O-C$ investigation. 
The first peculiar issue is the fact, that beside standard integer values of cycle numbers, we found $26$ maxima which are shifted by a half of a cycle with respect to these typical ones. 
They are shown in Fig.~\ref{fig:oc} as open circles. 
Another interesting matter is the jump of the phase by about $0.15$, observed somewhere around the cycle number $E~=~60$. Before and after the jump all the data points arrange as inclined straight lines. 

We separated the data into two data sets ($E~<~60$ and $E~>~60$) in order to check out if we observe only the phase jump itself, or rather a period change. 
We fitted linear functions to both data sets and obtained the following ephemerides for $E~<~60$ and $E~>~60$, respectively:
\begin{equation}
\rm{HJD}_{max} = 2455478.2865(7) + 0.066841(23) \times E
\end{equation}
\begin{equation}
\rm{HJD}_{max} = 2455478.306(2) + 0.066746(14) \times E
\end{equation}

The difference between these two periods is equal to $0.000095~\rm{d}$, which is four times greater than the accuracy of measurements. 
This result is quite ambiguous because it is almost bound together with the 3~sigma extent of the periods. 
Thus, a conclusion that at about the cycle $E~\approx~60$ there exists a change between two separate periods would probably be exaggerated. 

We claim that this is the effect of only a phase shift which is also often observed in other SU~UMa stars. 
An explanation of this feature is that at some point the main source of the superhump signal starts to be dominated by the bright spot rather than the flexing disc \citep{2011Wood}.
This is expected to happen at the moment of transition from the plateau phase to the decline of a light curve. 
However, in the case of this particular superoutburst of IX~Dra, the phase 
shift occurs before the transition, which is unusual.
It is unclear if this feature is specific to this particular superoutburst or it is an inherent characteristic of the system.
Previously published light curves of IX~Dra, which show double periodicity, are too complicated to identify secondary humps \citep{2004Olech}.

On the other hand, in another ER~UMa-type object, DI~UMa, secondary maxima are detected only on the very last days of superoutburst \citep[Fig.~7]{2009Rutkowski}.
It is also worth mentioning that the occurrence of such secondary maxima, shifted by a half of a cycle with respect to primary superhumps, is most probably a characteristic of only dwarf novae with short orbital periods. It was suggested by \cite{2007Rutkowski} that objects with longer orbital periods are less likely to show such behaviour.

\subsubsection{Normal outbursts and quiescence}

We performed the $O-C$ analysis also for the part of the light curve between both superoutbursts, covering a set of normal outbursts and a quiescent state. 
Although the time coverage here was much poorer, we were able to detect a number of clear humps.

Modulations of the light curve around maxima of normal outbursts are very weak. There is only a subtle flickering there, like in the case of typical SU~UMa stars.
Clear humps were identified only below the mean magnitude of outbursts, mainly very close to the minimum brightness of the star.

The power spectrum for this part of the light curve, shown in Sec.~\ref{sec:PSofNO}, is very complex in structure and shows several parallel peaks around the value $f~\approx~15~\rm{c/d}$.
This suggests only the fact that there is some periodicity present, but without any quantitative estimate.
That is why it was crucial to perform as optimal an $O-C$ analysis as possible, to verify the accurate value of a period observed during normal outbursts and quiescence. 
Thus, we precisely measured times of maxima of every single hump which we were able to catch, even those with a lower amplitude, or with a poor time coverage of observations. 

The $O-C$ diagram for all these points is rather noisy.
Nonetheless, a linear fit to these points allowed us to determine the ephemeris:
\begin{equation}
\rm{HJD}_{max} = 2455497.73(1) + 0.06639(3) \times E
\end{equation}
An interesting thing is the fact that the obtained period $0.06639(3)~\rm{d}$ is in agreement with the $P_{f_3}$, which in our opinion could be the orbital period of the system, and with the orbital period identified in \cite{2004Olech}.
However, this data set does not indicate any information about the stability of the obtained period. 

To check the stability of this period, we decided to perform one more improvement and restrict our $O-C$ analysis.
From the previous data set of all possible humps, we separated only 6 humps which are obviously clear and accurately covered by observations, thus are beyond any doubt.
For this sparse data set we performed another $O-C$ analysis.

From a linear fit to these points we obtained the following ephemeris:
\begin{equation}
\label{eq:bestHumps}
\rm{HJD}_{max} = 2455508.176(3) + 0.06646(2) \times E
\end{equation}

\begin{figure}
\begin{center}
\includegraphics[width=0.325\textwidth,angle=270]{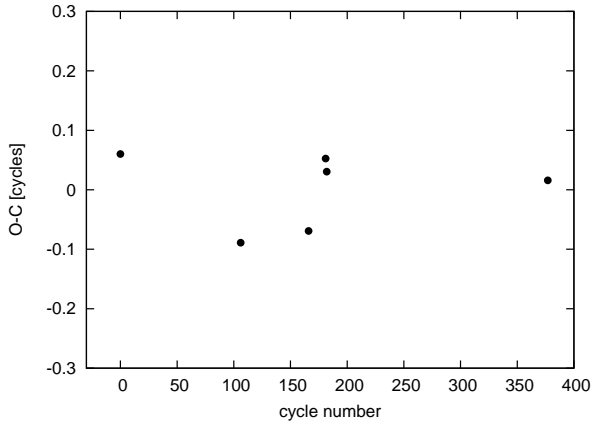}
\caption{The $O-C$ diagram for maxima of definite humps observed during normal outbursts and the quiescent state.}
\label{fig:oc_inB}
\end{center}
\end{figure}

The consequent $O-C$ diagram is shown in Fig.~\ref{fig:oc_inB}.

Times, errors, number of cycles ($E$), and $O-C$ values computed according to the linear ephemeris for each of these points are given in Tab.~\ref{tab:oc_inB}. 
\begin{table}
\centering
\caption{Cycle numbers $E$, times of maxima ($\rm{T_{max}}$, \textit{i.e.} $\rm{HJD}-2455000~[\rm{d}]$), and $O-C$ values (in units of days) 
of clear and definite humps observed during normal outbursts and the quiescent state.}
\begin{tabular}{c|c|c|c}
\hline
$E$		&	$\rm{T_{max}}$	&	error	&	$O-C$		\\
\hline
0		&	508.180		&	0.002	&	0.004		\\
106		&	515.215		&	0.003	&	-0.006		\\
166		&	519.204		&	0.004	&	-0.005		\\
181		&	520.209		&	0.003	&	0.003		\\
182		&	520.274		&	0.002	&	0.002		\\
377		&	533.233		&	0.002	&	0.001		\\
\hline
\end{tabular}
\label{tab:oc_inB}
\end{table}

Although based on this very limited data set, we obtained a result which is in agreement with the orbital period $P_{f_3}$ (Eq.~\ref{eq:bestHumps}).
What is more, the stability of this value is now confirmed on the basis of the $O-C$ diagram (Fig.~\ref{fig:oc_inB}).

\subsection{Periods' interpretation}
\label{sec:per_int}

All relevant periods emerging from our photometric data have been collected in Tab.~\ref{tab:periods}.

\begin{table}
\centering
\caption{Periods derived from our data. Values are expressed in days.}
\begin{tabular}{c|c|c}
\hline
	&	Superoutbursts	& Normal outbursts	\\
	&					& and quiescence	\\
\hline
power 		&	$0.06707(3)$	& $P_{f_1}~=~0.07148(4)$ \\
spectrum	&					& $P_{f_2}~=~0.06745(3)$ \\
analysis  	&					& $P_{f_3}~=~0.06641(3)$  \\
			&					& $P_{f_4}~=~0.06482(3)$  \\ \\

$O-C$ 		&	$0.066894(6)$	& $0.06646(2)$ \\
analysis	& & \\
\hline
\end{tabular}
\label{tab:periods}
\end{table}

Periods obtained during superoutbursts are consistent between both methods of analysis.
We detected only one clear periodicity, which is the superhump period, $P_{sh}$.
As a representative value, we consider here the corresponding mean value, namely, $P_{sh}~=~0.06698(4)~\rm{d}$.

The obtained $P_{sh}$ indicates that the value of the superhump period remains constant for this object over the last decade, see \citet{2001Ishioka} and \citet{2004Olech}.
Although, in contrast to the latter publication, we did not detect any orbital period modulation of the light curve throughout superoutbursts, which is surprising. 

The presence of an additional hump during superoutubust 
(see Fig.~\ref{fig:shevolution} and Fig.~\ref{fig:oc}) has been previously 
discussed, but it requires interpretation.
It appears just after four days since the beginning of the superoutburst (about the cycle number $E~\approx~60$) with the same period as typical superhumps ($P_{sh}$) but with a half-cycle phase shift.
At about $E~\approx~135$, amplitudes of these secondary humps become dominant relative to primary superhumps.
As a consequence, we observe a jump of the phase in the $O-C$ diagram (Fig.~\ref{fig:oc}).

An explanation of this feature could perhaps be the presence of an additional source of flux on the surface of the accretion disk of IX~Dra. It is worth mentioning here an alternative interpretation of superhumps which was presented by \cite{2007Smak} when analysing a SU~UMa-type object, Z~Cha. He showed that it is possible that the stream of matter overflows the disk of a system, and beside the "standard" hot spot, it causes additionally a second "modified" spot, presumably on the other side of the disk. Perhaps this could be the source of such double humps?
Such an explanation is tempting and it would imply a need of an improvement of the standard, commonly accepted tidal-resonance model of superhumps \citep{1988Whitehurst}.
\cite{2012Uemura} also showed a model with two flaring parts in the disk of SU~UMa-type star, which produce early superhumps with secondary humps. They also claim that the tidal effects alone cannot explain such a behaviour. 
On the other hand, \citet[and references therein]{2011Wood} mentioned an example of a "double humped" light curve with two approximately antiphased signals, and show a model with two distinct physical mechanisms which could cause positive superhumps, namely the viscous dissipation in the flexing disk and the time-variable viscous dissipation of the bright spot. 
Also, a proposal of a new interpretation of superhumps as they are caused due to enhanced dissipation of the kinetic energy of the stream was suggested by \cite{2009Smak}. This effect is a consequence of a modulated irradiation of the secondary component, and it would be in fact the most prominent for dwarf novae with very short orbital periods, \textit{i.e.} ER~UMa-type stars. 
Nevertheless, the interpretation of the observed additional humps in the case of IX~Dra is still open and needs further studies.

Let's get back to Tab.~\ref{tab:periods}.
When it comes to normal outbursts and the quiescent state, the power spectrum analysis gave us only the information that there exists some periodicity in the data. The $O-C$ analysis provided the quantitative result  of a constant value $P_{orb}~=~0.06646(2)~\rm{d}$, which is in perfect agreement with the value of orbital period of IX~Dra estimated by~\citet{2004Olech}, $P_{orb} = 0.06646(6)~\rm{d}$.

\begin{figure}
\begin{center}
\includegraphics[width=0.4\textwidth,angle=270]{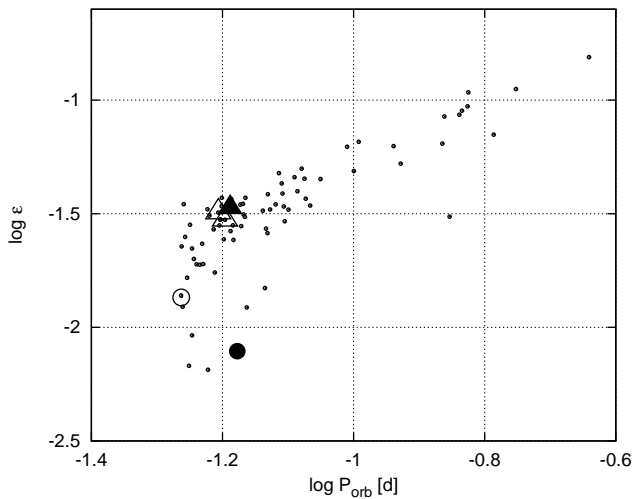}
\caption{
The relation between the orbital period-superhump period excess and the orbital period for dwarf novae stars. 
Small points represent known objects, from \citet{2011Olech}. 
The black dot corresponds to the position of IX~Dra obtained from $P_{orb}$ and $P_{sh}$.
The black triangle indicate the position of IX~Dra calculated from $P_{f_4}$ and $P_{sh}$.
Open symbols show positions of other ER~UMa-type objects: triangles stand for ER~UMa and V1159~Ori, and the circle for DI~UMa.}
\label{fig:epsilon}
\end{center}
\end{figure}

With the obtained values of $P_{sh}$ and $P_{orb}$ we can draw conclusions about the evolutionary status of IX~Dra. 
In Fig.~\ref{fig:epsilon} 
we show the position of the object in the standard $P_{orb}$~vs.~$\epsilon$ diagram for dwarf novae stars,
where $\epsilon$ is the superhump period excess defined as: 
\begin{equation}
\epsilon = \frac{\Delta P}{P_{orb}} = \frac{P_{sh}-P_{orb}}{P_{orb}}
\label{eq:epsilon}
\end{equation}
The period excess can be expressed in terms of the mass ratio of the binary system, $q~=~M_2/M_1$ \citep[and references therein]{Olech_QWSer}:
\begin{equation}
\epsilon \approx \frac{0.23q}{1+0.27q}
\label{eq:e-q}
\end{equation}
Thus, the $P_{orb}$~vs.~$\epsilon$ relation provides us an excellent plane to examine the evolution of the stars. 

When we apply the formula~\ref{eq:epsilon} to the final set of periods, $P_{sh}$ and $P_{orb}$, which were described above, we obtain the position of IX~Dra in the region of so-called \textit{period bouncers} (the dot in the Fig.~\ref{fig:epsilon}). 
In the same region there is located also another ER~UMa-type star, DI~UMa.

However, it was tempting to determine the location of IX~Dra for another prominent peak of the periodogram, which was presented in Sec.~\ref{sec:PSofNO}, namely $P_{f_4}$.
Although we did not confirm this value by the $O-C$ analysis, its amplitude in the power spectrum was slightly higher than the one of $f_3$. 
The resulting position of IX~Dra in the $P_{orb}$~vs.~$\epsilon$ diagram was surprising.
In the case of $P_{f_4}$, the object is located in the area where typical SU~UMa stars are found before they reach the minimum period during their evolution (the triangle in the Fig.~\ref{fig:epsilon}). 
In the same region there are located also other two of five known ER~UMa-type stars, ER~UMa and V1159~Ori.

Previous studies show that IX~Dra is located in the \textit{period bouncers} region in this diagram \citep{2004Olech}.
This is supported by our observations, but we cannot exclude the other possibility, related with the $P_{f_4}$, which was discussed here.

\section{ER UMa stars}
\label{sec-eruma}

After 17 years since the distinction from typical SU~UMa stars \citep{1995Kato, 1995Robertson},
the ER~UMa-type of dwarf novae is still very poorly understood. 
Until now there are discovered only five stars of this kind.
Thus, studying each one of them is extremely important in order to draw conclusions regarding this class of particularly active dwarf novae.
Because of the fact that the behaviour of ER~UMa-type stars is so untypical, in many cross-sectional statistical studies of dwarf novae they are often simply discounted, see for instance \citet{2003KatoHODel} and \citet{2011Patterson}. 
That is why particular attention shall be paid to investigate this class of stars separately.

We performed a literature survey of ER~UMa-type dwarf novae and, together with our new results, we present a new set of basic statistics. See Tab.~\ref{tab:eruma}.

\begin{table*}
\centering
\begin{minipage}{175mm}
\caption{
Properties of known ER~UMa-type stars. 
$P_{orb}$ and $P_{sh}$ are the orbital and superhump periods.
$P_{sc}$ and $P_c$ are the supercycle and cycle lengths, 
$D_{s}$ and $A_{s}$ are the duration and amplitude of superoutbursts, 
and $D_{n}$ with $A_{n}$ are the same for normal outbursts. 
$\epsilon$ is the period excess, and $\epsilon_-$ is the period deficit in the case of a presence of negative superhumps.
\textit{Reg.} stands for \textit{region} in the $P_{orb}$~vs.~$\epsilon$ diagram (see Fig.~\ref{fig:epsilon}):
the black triangle symbol indicates a position in the region of typical SU~UMa stars before they reach the minimum period during their evolution, and the dot corresponds to a position of an object in the area of \textit{period bouncers}.
The last two rows show both possible sets of parameters of IX~Dra, depending on the choice of the orbital period (see Sec.~\ref{sec:per_int} for details).
}
\begin{tabular}{@{}lllllllllllll@{}}
\hline
Object  	& $P_{orb}$ 		&  $P_{sh}$ 	& $P_{sc}$ 		& $P_c$ 	& $D_{s}$ 	& $A_{s}$ 	& $D_{n}$ 	& $A_{n}$ 	& $\epsilon$ 	&	$\epsilon_-$	&	$Reg.$ 				& Ref.	 	\\
	    	& [d]  				& [d]  			& [d]   		& [d]  		& [d]   	& [mag]  	& [d]   	& [mag]  	& $\%$  		&	$\%$			&						&			\\
\hline
ER~UMa  	& 0.06366(3) 		&  0.066862798	&  43-45		& 4-6 		& 20		& 3			& 3-4		& 2-2.5		& 2.1  			&	$-7.5$		&	$\blacktriangle$	&	\footnote{\label{note1}\cite{2009Wood}}, \footnote{\label{note2}\cite{1997Thorstensen}}, \footnote[3]{\cite{1999Gao}}, \footnote[4]{\cite{2006Zhao}} \\
V1159~Ori 	& 0.06217801(13)	&  0.064167(40)	&  44.6-53.3	& 4   		& 20		& 2.5		& 3-5		& 2			& 3.2  			&	$-7.6$		&	$\blacktriangle$	&	\footnoteref{note1}, \footnoteref{note2}, \footnote[5]{\cite{1995Patterson}}, \footnote[6]{\cite{2001Kato}}  \\
RZ~LMi  	& ---	 			&  0.059396(4)	&  19.07(4)		& 4.027(3)	& 10		& 2.6		& 4-5		& 2.1		& ---  			&	---			&	?					&	\footnote[7]{\cite{2008Olech}}  \\
DI~UMa  	& 0.054579(6)		& 0.055318(11) 	&  31.45(30)	& ?   		& 12		& 3.3		& ?			& 2.4		& 1.4	 		&	---			&	$\bullet$			&	\footnote[8]{\cite{2009Rutkowski}}  \\
IX~Dra (1) 	& 0.06646(2)		& 0.066982(36)  &  58.5   		& 3.1-4.1	& 15  		& 3.3 		& 4 		& 2.8 		& 0.8   		&	---			&	$\bullet$			&	this work\\
IX~Dra (2)	& 0.06481(5)		& 0.066982(36)  &  58.5   		& 3.1-4.1 	& 15  		& 3.3 		& 4 		& 2.8 		& 3.4	   		&	---			&	$\blacktriangle$	&   this work\\ 
\hline
\end{tabular}
\label{tab:eruma}
\end{minipage}
\end{table*}

The first evident conclusion from Tab.~\ref{tab:eruma} is the fact that the class of ER~UMa-type stars is not homogeneous. 
It is possible to divide them into two separate groups of objects with similar characteristics, which determine their evolutionary status (on the assumption that these extremely active stars obey the standard relation for typical dwarf novae stars, see Eq.~\ref{eq:epsilon} and \ref{eq:e-q}).

ER~UMa and V1159~Ori are very alike and they form the first of these groups (group I).
With the highest period excesses, they are located in the same region of the $P_{orb}$~vs.~$\epsilon$ diagram, unlike the rest of the objects. They seem to have normal secondaries and evolve towards the minimum orbital period.
They are the only two members of the ER~UMa class which show occasionally negative superhumps \citep{1995Patterson, 1999Gao}.
Durations of their superoutbursts ($D_{s}=20\rm~d$) are significantly longer than of the other objects ($D_{s}=10-15\rm~d$). 

On the other hand, IX~Dra with the first version of parameters (see Tab.~\ref{tab:eruma}) is very similar to DI~UMa, especially when it comes to the period excess and location in the $P_{orb}$~vs.~$\epsilon$ diagram. Thus, they have the same evolutionary status and we believe that they are period bouncers, \textit{i.e.} more evolved objects with degenerate brown dwarfs as secondaries, evolving now toward longer periods (group II).
The second set of parameters of IX~Dra is possible, but less likely. In this case the object would be a much younger star still evolving towards the period minimum, so it would be more similar to ER~UMa and V1159~Ori. 

The last object, RZ~LMi, is the least known one. 
Its outbursts parameters are similar to those of other ER~UMa stars, but RZ~LMi is definitely the most active among them ($P_{sc}$=19.07(4)~d). 
This object shows permanent superhumps, both in superoutbursts and quiescence. An interesting thing is the fact that phases of its superhumps are coherent even between different superoutbursts \citep{2008Olech}.
Unfortunately, there is no orbital period reported for this object, so its evolutionary status remains unknown.
The superhump period of RZ~LMi is very similar to that of DI~UMa. 
Perhaps its $P_{sc}$ will evolve towards longer values over next years (like we find for IX~Dra), and RZ~LMi will resemble objects of the group II.

Despite the possible division of all these objects into groups I and II, there are a number of similarities between them. 
They all have extremely short supercycles, and relatively low amplitudes of outbursts.
Also, V1159 Ori shows a double-peaked structure within its oscillations during superoutburst, which peak near the end of the superoutburst \citep{1995Patterson}, resembling IX~Dra and its possible additional source of flux on the surface of the accretion disk. 

In general, we understand the overall behaviour of ER~UMa-type of stars. \cite{1995Osaki} tried to reproduce the light curve of ER~UMa, based on the standard thermal-tidal instability model. He increased the mass transfer rate by a factor of 10 compared to that expected in the standard cataclysmic variable evolution theory to obtain an appropriate result.
With this model it is possible to reproduce normal and superoutbursts, together with superhump phenomenon during superoutbursts, which are defined as a beat between the precession of the accretion disk and the orbital motion of the binary. 

However, it is difficult to reconcile such an explanation with period bouncers.
From the evolution theory we know that it is unexpected and unusual for the most evolved systems to have such a high mass accretion. 
A typical dwarf nova evolves from long ($\sim~10$~h) to short ($\sim~70$~min) orbital period due to angular momentum loss. This stage of evolution corresponds to the mass transfer rate decrease from $10^{-8}$ to $10^{-10}~\rm{M_\odot/yr}$.
After the star reaches the minimal period, the secondary becomes a degenerate brown dwarf and the system starts to evolve again to longer orbital periods. At this stage the mass transfer rate drops to about $10^{-12}~\rm{M_\odot/yr}$ and the object becomes inactive \citep{2011Knigge}. 
Thus, the older the system, the less active it should be. 
IX~Dra and DI~UMa as period bouncers with brown dwarf donors do not fit this explanation. 

Nonetheless, the behaviour of ER~UMa-type stars seems to be very complex. 
We need some improvement in the existing models to explain also the complicated shape of the oscillations in the light curves. 
\cite{1999Gao} claims that "it is more likely that superhumps occasionally exist at essentially all phases of the eruption cycles of ER UMa stars", not only during superoutbursts, which is in fact observed. 
It was suggested by \cite{Smak2004a,Smak2004b} that some problems with the TTI model for superoutbursts might be solved by a hybrid model which could combine the TTI model with the irradiation-enhanced mass-transfer effects. 
This could improve modelling of the observed superhump amplitudes. Effects of the irradiation-enhanced mass outflow are the most significant in systems with short orbital periods, so it is an important, probably crucial, issue in modelling ER~UMa-type stars.

Another important problem is the lack of high-quality spectra for these 
objects. They are certainly needed to obtain precise orbital periods, but also to verify the nature of secondary stars in these systems.
\cite{2007Ishioka} showed that spectra of IX Dra exhibit no features of the secondary star. It was impossible to distinguish between models of M2 dwarf from others. They suggested that it is possible "that the masses of the secondary stars in the ER~UMa systems are similar, but their temperatures are higher than those of normal SU~UMa stars". This would have a significant influence on the evolutionary status of these objects, and would verify the application of the standard relation of typical dwarf novae stars (Eq.~\ref{eq:epsilon}) for ER~UMa-type of stars. 

\section{Conclusions}
\label{sec-sum}

We report results of a new world-wide observing campaign of IX~Draconis, together with our analysis of the power spectra and the \textit{O-C} diagrams.
We discovered that the supercycle length $P_{sc}$ has been increasing over the last twenty years, and it is now equal to $58.5\pm0.5~\rm{d}$. 
The rate of the increase of the supercycle length is $\dot{P} = 1.8 \times 10^{-3}$.
Normal outbursts appeared irregular during our observations, with the interval between two successive normal outbursts between $3.1 - 4.1~\rm{d}$. 
We detected a double-peaked structure of superhumps during superoutburst, with the secondary maximum peaking near the end of the superoutburst.  
The mean superhump period observed during superoutbursts is $P_{sh} = 0.066982(36)~\rm{d}$.
We found two possible values of the orbital period: the first one, $0.06641(3)~\rm{d}$, which is in agreement with previous studies and our $O-C$ analysis ($0.06646(2)~\rm{d}$), and the second one, $0.06482(3)~\rm{d}$, which is less likely.
The evolutionary status of the object changes dramatically, depending on the choice between these two values. 

To conclude, we updated available information on ER~UMa-type stars and present a new set of their basic statistics. 
We emphasise the lack of spectroscopic observations for these objects, which could provide many answers regarding this class of stars.

\section*{Acknowledgements}
We are very grateful to Professor J\'ozef Smak for his comments which helped to improve the text of the paper, to an anonymous referee for his/her thoughtful comments, and to Alexis Smith for carefully reading this manuscript. 

We gratefully acknowledge the generous allocation of time at the Warsaw Observatory 0.6-m telescope and at the Tzec Maun Observatory.
We acknowledge with thanks the variable star observations from the \textit{AAVSO} International Database contributed by observers worldwide and used in this research.
This research has made use of the Simbad database, operated at CDS, Strasbourg, France.

The project was supported by Polish National Science Center grant awarded by decision number DEC-2012/04/S/ST9/00021 to AR, and Polish National Science Center grant awarded by decision number DEC-2011/03/N/ST9/03289 to MOH.

\bibliographystyle{mn2e} 
\bibliography{ixdra_bib}

\bsp

\end{document}